\documentclass[prb,twocolumn,superscriptaddress,preprintnumbers,amsmath,amssymb,floatfix]{revtex4}
\usepackage{graphicx}

\newcommand{\beq}{\begin{equation}}
\newcommand{\eeq}{\end{equation}}

\begin{document} 
\title{
Titanium Nitride - a correlated metal at the threshold of a
Mott transition
}
\author{H. Allmaier}
\email[]{hannes.allmaier@itp.tugraz.at}
\affiliation{Institute of Theoretical and Computational Physics, Graz University of Technology,
A-8010 Graz, Austria}
\author{L. Chioncel}
\affiliation{Institute of Theoretical Physics, Graz University of Technology,
A-8010 Graz, Austria}
\affiliation{Faculty of Science, University of Oradea, RO-47800, Romania}
\author{E. Arrigoni}
\affiliation{Institute of Theoretical Physics, Graz University of Technology,
A-8010 Graz, Austria}

\begin{abstract} 
We investigate electron correlation effects in stoichiometric Titanium Nitride (TiN) 
using a combination of electronic structure and many-body calculations. In a first 
step, the Nth-order muffin tin orbital technique is used to obtain parameters for 
the low-energy Hamiltonian in the {\it Ti-d($t_{2g}$)}-band manifold. The 
Coulomb-interaction $U$ and the Hund's rule exchange parameter $J$ are estimated
using a constrained Local-Density-Approximation  calculation. Finally,  the 
many-body problem is solved within the framework of the Variational Cluster Approach. 
Comparison of our calculations with different spectroscopy results stresses the 
importance of electronic correlation in this material. In particular, our results 
naturally explain a suppression of the TiN density of states at the Fermi level 
(pseudogap) in terms of the proximity to a Mott metal-insulator transition. 
%
%
\end{abstract}

\maketitle 

\section{Introduction.}
Transition metal-nitrides have been studied for several decades due to their appealing 
properties, such as ultra-hardness, high melting point and high 
Curie-temperature. This combination of physical and chemical characteristics 
makes them particularly suitable for coating applications. These materials 
exhibit metallic conductivity and some of them even show superconductivity, as for example the Nb 
based carbonitride \cite{sc.wi.85} which has a transition temperature of 18K.
Recent low-temperature transport properties of thin TiN superconducting films
~\cite{vi.ba.08} 
 show a disorder-driven 
transition from a superconductor to an insulating phase in which superconducting 
correlations persist. These experiments performed on homogeneously disordered
TiN films clearly demonstrate the important role of electronic correlations.
For these reasons, one could expect that signatures of many-body 
effects might be also present in bulk, stoichiometric TiN. 

Electronic properties of bulk transition metal nitrides have been
investigated using X-ray Photoelectron Spectroscopy (XPS)~\cite{so.ab.97,po.ro.83}, 
Ultraviolet Photoemission Spectroscopy (UPS)~\cite{jo.st.80}, X-ray emission \cite{gu.ku.77}, Bremsstrahlung-Isochromat Spectroscopy 
(BIS) \cite{ri.wo.82}, and Electron-Energy-Loss Spectroscopy (EELS)~\cite{sc.sh.81,sc.sh.81b}. 
A number of optical reflectivity measurements have also been carried out~\cite{ly.ol.80,al.fo.75}. 
On the theoretical side, a large number of band structure calculations using density functional 
theory (DFT) - mostly within the local-density-approximation (LDA) - 
is present in the literature~\cite{neck.83,ma.we.86,pa.pi.85,so.ab.97}.  LDA results
describe bonding in terms of (i) a {\it metallic} contribution giving a finite density 
of states at the Fermi level, (ii) an {\it ionic} contribution caused by the charge transfer 
from the metal to the non-metal atom, and (iii) a {\it covalent} contribution due to the 
interaction between the non-metal {\it p} and the metal {\it d} valence states in addition 
to the (iv) {\it metal-metal} interactions. In the case of TiN, it is believed that bonding 
is mostly covalent in origin. However, while the LDA results for the occupied part 
of the density of states (DOS) show a good agreement with spectroscopic data 
{\em at high binding energies}, they fail to describe correctly the energy range 
in the vicinity of the
Fermi energy (see, e. g., Fig.~\ref{doslda})
where predominantly Ti-$3d$ states are present~\cite{zh.gu.88,jh.lo.01,pa.sc.00}.
In particular, a suppression of the 
 XPS spectrum
over a range of energy down to $\approx 1 eV$ below the Fermi energy, could be
explained so far only by means of an artificial broadening, whose
parameters are optimized to fit the experimental data.
In this paper, we present an alternative view and argue
that this suppression can be naturally
explained in terms of correlation effects, and, specifically, by the
proximity to a Mott metal-insulator transition.

Recent developments in the methodology of  electronic-structure calculations allow 
to go beyond the LDA and include {\it electronic correlation} effects, which are
particularly important for electrons in $d$ or $f$ orbitals.
The simplest mean-field extension of LDA is the LDA+U 
approach~\cite{an.ar.97}. 
However, as we show
 below,  even such  extension is not sufficient to
improve the agreement with experiments in TiN.
For this reason, it is necessary to go beyond the mean-field approach and to
  appropriately deal with {\it dynamical } correlation effects.
In this paper, we carry out this task by means of a cluster-perturbative method, the
Variational Cluster Approach (VCA)~\cite{po.ai.03,da.ai.04}, which we combine
with LDA in order to obtain the appropriate model parameters.

This paper is organized as follows: the results of the electronic-structure 
calculations  within the framework of the linear muffin-tin orbital method (LMTO)\cite{an.je.84}
 at the LDA level are presented in section~\ref{sec:nmto}.
Comparison of LDA and LDA+U results with experiments are also
discussed in that section.
In Sec.~\ref{abin} we describe the {\it ab-initio} construction of the many-body model 
Hamiltonian. Specifically, the uncorrelated part of the Hamiltonian for excitations 
in the vicinity of the Fermi level is obtained from the so-called downfolding
technique~\cite{an.sa.00,zu.je.05} within the Nth-order muffin tin orbital 
(NMTO) method. The {\it interaction} part is then estimated by the constrained-LDA method.
In Sec.~\ref{vca}, we give a short summary of the VCA approach.
We present and discuss our results in Sec.~\ref{resu}. In particular,
in Sec.~\ref{dos},
we evaluate the density of states within VCA and compare it with
experiments, and discuss the results in the framework of previous
calculations.
In Sec.~\ref{spec} we discuss $k$-dependent spectral properties,
namely the spectral function and the self-energy,
 and evaluate the effective electron mass which we also compare to experiment.
Finally, we summarize our results in Sec.~\ref{summ}.

\section{Electronic structure calculations for TiN}
\label{sec:nmto}
\begin{figure}[h]
\includegraphics[width=0.90\columnwidth]{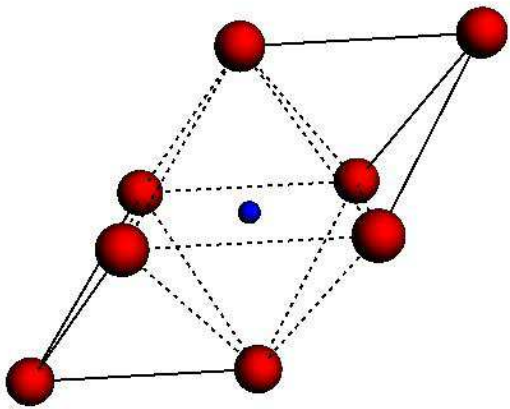}
\caption{(color online) Octahedral environment formed by the Ti-atoms
(large, red spheres) around the N atom (small, blue sphere).
The unit cell is represented by the black lines; the subset of Ti-atoms connected
by dashed lines shows the cluster geometry that was used as reference system for the 
VCA-calculation (see text).}
\label{tin_fig}
\end{figure}

TiN crystallizes in the rock-salt (B1) structure where Ti and N atoms
are sitting 
on interpenetrating face-centered cubic lattices originating  at
 $(0,0,0)$, and  
at  $(\frac 1 2,  \frac 1 2, \frac 1 2)$ in units of  the lattice
parameter  $a=7.65 a_0$ ($a_0=$ Bohr radius), respectively. 
 Each Ti (N) atom  has six 
 N (Ti) nearest neighbors in an octahedral geometry. 
Fig.~\ref{tin_fig} shows the conventional unit cell containing 
Ti (large spheres, red) and 
N (small sphere, blue)
atoms.
The electronic configuration of the outer shell of Ti is $4s^2\ 3d^2$.
Therefore, following from the formal oxidation $Ti^{3+}$, there is a single 
Ti electron in the valence band.
 Due to the 
octahedral symmetry, the Ti-$d$-orbitals are split into the three $t_{2g}$ 
orbitals at lower energy, and two $e_g$ orbitals at higher energy.

\begin{figure}[h]
\includegraphics[width=0.99\columnwidth]{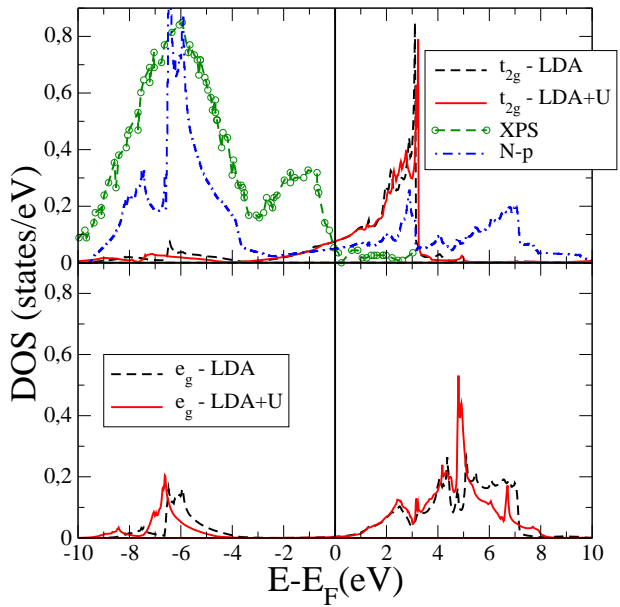}
\caption{(color online) 
Orbital-resolved density of states 
for Ti obtained by LDA and LDA+U. 
The upper (lower) panel shows the
$t_{2g}$ ($e_{g}$) contribution.
The  upper panel also displays the N-$p$ density of states 
and the 
experimental XPS spectrum~\cite{so.ab.97} 
(Experimental data reproduced with kind permission of the authors).
}
\label{doslda}
\end{figure}

The LDA band structure of TiN was computed with the LDA-LMTO (ASA) code~\cite{an.je.84}
which uses the basis of linearized muffin-tin orbitals in the atomic sphere approximation. 
Numerically,
two empty spheres per unit cell have to be introduced at the positions $(1/4, 1/4, 1/4)$ 
and $(3/4, 3/4, 3/4)$, 
due to the atomic sphere approximation (ASA) used in the calculation.
%
Results are shown in Fig.~\ref{doslda}. In the density of states, the N-$s$ orbital (not shown) 
form the lowest valence band widely separated from the other valence bands. 
At higher energies, one finds the set of bands formed by N-$p$ orbitals situated between -10 
and -4eV. Finally, the last five bands mainly consist of Ti-d orbitals. The DOS around the
Fermi level is dominated by Ti-$t_{2g}$ bands, while $e_{g}$ bands remain above the Fermi 
level and are empty. $t_{2g}$ and $e_g$ bands overlap in the unoccupied part
of the spectrum. On the other hand, the hybridization of Ti-d orbitals with N-$p$ orbitals 
below the Fermi level is at the origin of the covalent bonding.
Concerning the $p-d$ hybridization, the energetically higher $d$-bands form the 
anti-bonding states while the lower $p$-bands form the bonding
states. The position of the Fermi level ($E_F$) is determined by the number of 
valence electrons per unit cell being equal to 9.
%

In order to take into account correlation effects on a mean field-level, we carried out an 
LDA+U calculation. Here, we used
values of $U=10$eV and $J=1.3$eV, as obtained from the constrained LDA
calculations~\cite{an.za.91,an.gu.91}.
As one can see there are no significant differences between the LDA
and the LDA+U  results despite the large value of $U$. 
However, a comparison of these calculations 
with valence band XPS spectra, which provide
a measure of the total density of occupied states as a function of energy, 
shows that neither LDA nor LDA+U results are appropriate  
to describe the experimental spectra in the vicinity of the Fermi
energy unless one introduces ad-hoc broadening terms.
Specifically, while  the position of the N-$p$ bands
obtained by LDA and LDA+U is in reasonable agreement 
with  the XPS measurements (see Fig~\ref{doslda}), 
both methods fail to reproduce the behavior of the DOS within a range
of $\approx 2$ eV below the Fermi energy.
 In particular, the XPS spectra show a local maximum 
at energies of $\approx -1$eV, followed by a ``pseudogap''-like suppression
at the Fermi energy.
The LDA+U results do not change qualitatively when decreasing $U$
down to $U\approx 8$, which, as argued below, is more appropriated
for this material. 

These results suggest that static inclusion of correlations is not
sufficient to explain the DOS near the Fermi energy.
For this reason, we have taken into account dynamical correlation
effects beyond the LDA results by means of
the VCA, a method appropriate to treat correlated systems.
This approach  builds up on the exact
diagonalization of an Hamiltonian on a finite cluster combined with an
appropriate extension to the infinite-lattice limit.
However, in order to limit the size of the Hilbert space, it is necessary to
use
an effective Hamiltonian describing a small number of correlated
effective orbitals
 per lattice site in the close vicinity of the Fermi level.
From Fig.~\ref{doslda}, we conclude that the minimal model has to be  
restricted to Ti ($t_{2g}$) bands, while higher 
 Ti ($e_{g}$) bands are unoccupied and can be neglected.

\subsection{Ab-initio construction of the model Hamiltonian}
\label{abin}

In order to construct the effective low-energy Hamiltonian to use in
our VCA calculation, 
 we employed the Nth order muffin-tin-orbitals scheme
within the same LMTO-ASA basis set.
The NMTO method~\cite{an.sa.00,zu.je.05} can be used to generate 
truly minimal basis sets with a massive 
downfolding technique. 
This reduced basis set reproduces
the bands obtained with the full basis set to great accuracy and
can thus be used as non-interacting part of the many-body Hamiltonian.
According to the discussion in Sec.~\ref{sec:nmto}, 
the minimal basis set is obtained by downfolding to the
 {\it Ti-d($t_{2g}$)} manifold.
The truly minimal set of symmetrically orthonormalized NMTOs is a set of Wannier 
functions.
 In the construction of this set, the active channels are forced 
to be maximally localized onto the eigenchannel ${\bf R}lm$ ($\bf R=$
atomic position, $l,m=$ angular momentum quantum numbers), which makes the
basis set strongly localized and suitable for a real-space Hamiltonian.

\begin{figure}[h]
\begin{center}
\includegraphics[angle=270,width=1.\columnwidth]{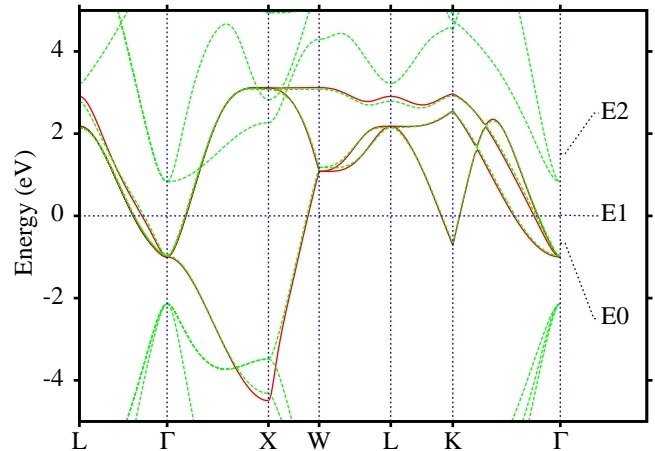}
\end{center}
\caption{(color online) Band-structure of TiN 
calculated for 
a path in the Brillouin zone (BZ) going 
 from the L-point $(0.5,0.5,0.5)$ trough the $\Gamma$  $(0,0,0)$, $X$
 $(0,1,0)$, $W$ $(0.5,1,0)$, $L$ $(0.5,0.5,0.5)$, $K$ $(0,0.75,0.75)$
 points and ending at the $\Gamma$ point $(0,0,0)$.
The bands are obtained from the 
LDA calculation (dashed/green line) 
for the complete orbital basis set. The NMTO bands (full/red line) are 
obtained after downfolding to the Ti-$3d$($t_{2g}$) orbitals.}
\label{NMTO_bnds}
\end{figure}

In this way, 
the  non-interacting part of the effective Hamiltonian 
is confined to a reduced set of orbitals and to a 
reduced energy window. 
The NMTO downfolded bands are obtained by expanding around a small number of
energy points on which the LDA bands are reproduced exactly.
To optimize the overall agreement with the energy bands, 
 we chose the following expansion points
$E_\nu-E_F=1.512$eV, $0.016$eV, and  $-0.664$eV. 
Results are quite stable for choices of the $E_\nu$ around these values.
In Fig.~\ref{NMTO_bnds}
we show the eigenvalues of the effective Hamiltonian along some high-symmetry directions
in comparison with the bands obtained from the full orbital basis. 
From this figure it is clear that the Ti-d($t_{2g}$)  manifold is well described 
by the non-interacting part of the effective Hamiltonian.

Fourier-transformation of the orthonormalized NMTO Hamiltonian, 
$H^{\rm LDA}({\bf k})$, yields on-site energies and hopping integrals
(${\bf R}\equiv(x,y,z)$),
\begin{equation}
\left\langle \chi _{%
{\bf R'}\,m^{\prime }}^{\perp }\left\vert H
^{LDA}-\varepsilon _{F}\right\vert \chi _{{\bf R} \,m}^{\perp}\right\rangle 
\equiv t_{m^{\prime },m}^{\bf{R'-R}\rm} ,
\end{equation}%
in a Wannier representation, where the NMTO Wannier functions
$\left\vert \chi _{\mathbf{R} \,m}^{\perp}\right\rangle$ are orthonormal.
In the restricted NMTO basis, $m$ is labeled by the three 
$t_{2g}$ orbitals $m=xy,yz,zx$ 
(we use this order for the definition of the matrix elements below).
In this basis, the on-site matrix elements  $t_{m',m}^{ 000}$ are diagonal and
independent of $m$. The precise value of these terms, 
as well as the
so-called 
double-counting correction, are not important, as they can
be absorbed into the chemical potential.
The directional hopping matrix elements up to second nearest-neighbor are given by
\begin{eqnarray}
\label{mat_ham}
t_{m',m}^{ \frac{1}{2}\frac{1}{2}0} &= \left(
\begin{array}{rrr}
 -0.66 &    0 &   0    \\
  0    & 0.15 & -0.1  \\
  0    & -0.1 & 0.15   
\end{array}
\right) \ \text{} , \nonumber   \\
t_{m',m}^{ 100} &= \left(
\begin{array}{rrr}
-0.23 &    0  &    0   \\
   0  &  0.01 &    0   \\
   0  &    0  & -0.23  
\end{array}
\right) \ \text{} \nonumber ,
\end{eqnarray}
in units of eV.
Only one representative hopping integral 
is shown for each class. 
Other hopping terms can be derived 
from proper unitary transformation using crystal symmetry. 
See, e.g.,
 Ref.~\onlinecite{pa.ya.05} for details.

Further hoppings are taken into account up to 
a range of $r=1.1 a$.
Neglected hoppings
are at least  by a factor 
40
smaller than the largest nearest-neighbor hopping.

In order to make sure that neglecting $e_g$ orbitals is safe, 
we have also carried out a LDA+VCA calculation (for smaller clusters)
 using an Hamiltonian downfolded to {\it all $5$}
 Ti-d bands. We
 have verified that the occupation of the $e_g$ bands is less
than
$10^{-3}$ per Ti-atom,
so that the corresponding Wannier functions can be safely neglected in
our calculation.
Notice, however, that, due to hybridization, $t_{2g}$ Wannier
functions also have a certain amount of $e_g$ character.

The non-interacting part of the effective Hamiltonian for TiN, thus, has the form
\begin{equation}\label{h0}
H_0 = \sum_{ {\bf R'},{\bf R}, \{m',m\},\sigma } t_{m', m}^{\bf R'-R} c^{\dag}_{{\bf R'} m'\sigma} c_{{\bf R} m\sigma } .
\end{equation}
To take into account correlation effects, we add the usual interaction term
\begin{eqnarray}
\label{hi}
H_I &=& \sum_{{ {\bf R}, m }}  U n_{ {\bf R}m\uparrow} n_{ {\bf R}m\downarrow}  \\ \nonumber
    &+& \sum_{{ {\bf R}, m < m',\sigma,\sigma'}} (U'-J\delta_{\sigma,\sigma'}) n_{ {\bf R}m\sigma}  n_{ {\bf R}m'\sigma'} \\ \nonumber
    &+& \sum_{{ {\bf R}, m < m'}} J c^\dag_{ {\bf R}m'\uparrow}c^\dag_{ {\bf R}m\downarrow}c_{ {\bf R}m'\downarrow}c_{ {\bf R}m\uparrow}+\text{h.c.} \\ \nonumber
    &+& \sum_{{ {\bf R}, m < m'}} J c^\dag_{ {\bf R}m'\uparrow}c^\dag_{ {\bf R}m'\downarrow}c_{ {\bf R}m\downarrow}c_{ {\bf R}m\uparrow}+\text{h.c.} .
\end{eqnarray}
In Eq.~\ref{hi},
$c_{{\bf R}m\sigma}$ $(c^\dag_{{\bf R}m\sigma})$ are the usual fermionic annihilation (creation) operators acting on an electron 
with spin $\sigma$ at site {\bf R} in the orbital $m$ and $n_{ {\bf
    R}m\sigma}=c^\dag_{{\bf R}m\sigma}c_{{\bf R}m\sigma}$. $U$ denotes
the Coulomb-interaction for two electrons in the same orbital with
anti-parallel spin. If they
are located 
on
two different orbitals, the interaction is reduced to
$U'=U-2J$. 
$J$ is the Hund's rule exchange constant and h.c. denotes the hermitian conjugate.

We have estimated the value of the Coulomb-interaction parameter
by means of constrained LDA, \cite{an.za.91,an.gu.91} whereby 
occupancies on all $d$ orbitals have been fixed. 
Since the  $e_g$ effective 
orbitals are essentially empty, as discussed above, they do not contribute
to screening and can be safely neglected.
The constrained calculation yields a value of $U\approx 10$eV and 
$J\approx 1.3$eV. However, as it was shown by Aryasetian et al~\cite{ar.ka.06}, 
constrained LDA gives larger values for $U$ 
compared to other 
methods based on the evaluation of the screened Coulomb interaction 
within the Random Phase Approximation~\cite{ar.im.04}.
Therefore, we have also investigated smaller $U$ values.
%
We will show below
that results change drastically around $U\approx 9$ eV, where a
Mott-insulator transition takes place.

As correlations on a mean-field level are already 
included in LDA, one should in principle subtract the long discussed 
double-counting~\cite{cz.sa.94,pe.ma.03}
correction. However, since $t_{2g}$ orbitals are degenerate, this
correction is simply a constant that can be absorbed
in the chemical potential.

\subsection{Variational Cluster Approach}
\label{vca}

To solve the many-body Hamiltonian \eqref{h0}+\eqref{hi}
 we employ the Variational Cluster Approach~\cite{po.ai.03,da.ai.04}.
This method is an
extension of Cluster Perturbation Theory (CPT)~\cite{gr.va.93,se.pe.00,ov.sa.89},
 in which the original lattice is
divided into a set of disconnected clusters and the inter-cluster hopping
terms are treated perturbatively.  
VCA additionally includes ``virtual''
single-particle terms to the cluster 
Hamiltonian, 
yielding a
so-called reference system, and then subtracts these terms perturbatively. 
The ``optimal'' value for
these variational parameters is determined in the framework of the Self-energy
Functional Approach (SFA)~\cite{pott.03,pott.03.se}, by requiring that the
 SFA grand-canonical potential $\Omega$
is stationary within this set of variational parameters. 
Since TiN is paramagnetic, we only include the chemical potential of the cluster as
 a variational parameter.
The latter is necessary in order to obtain a thermodynamically consistent
particle density~\cite{ai.ar.05,ai.ar.06}.
In this paper, we use a new method, described in Ref.~\onlinecite{lu.ar.09u},
to carry out the sum over Matsubara frequencies required in the
evaluation of $\Omega$, whereby an integral over a contour 
lying a distance $\Delta$
from the real axis is carried out. The crucial point is that the
contour integral is exact for any (even large) $\Delta$.

As a reference system we adopt 
the minimal cluster that contains the full lattice symmetry. This
consists of all $6$ Ti sites lying on the corners of an octahedron, as
shown in 
Fig.~\ref{tin_fig}.
Larger clusters are at present not feasible within our variational procedure.
As in the case of cluster-DMFT (which is currently not feasible for a 6-site
cluster with three orbitals each), the appropriate periodisation is a
crucial issue~\cite{bi.pa.04,sene.08u}.
Here, we choose to periodise the Green's function, as it is well known
that the self-energy
periodisation gives unphysical results in the vicinity of an
insulating phase.

\section{Results}
\label{resu}

\subsection{Density of states}
\label{dos}

\begin{figure}[h]
\includegraphics[width=0.95\columnwidth]{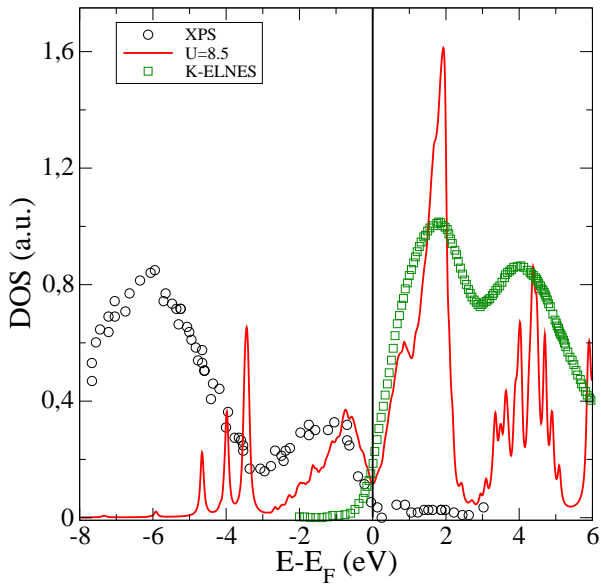}
\caption{(color online)
TiN density of states calculated within 
 LDA+VCA for $U=8.5 $eV and 
$J= 1.3$
 compared with measured 
XPS~\cite{so.ab.97} and K-ELNES~\cite{pa.sc.00} spectra
(Experimental data reproduced with kind permission from the authors).
}
\label{doscomp}
\end{figure}

As discussed in Sec.~\ref{sec:nmto}, the low-energy XPS~\cite{so.ab.97}
 spectrum is
characterized by a peak at about $-1$ eV 
followed by a ``pseudogap'', i. e. a suppression of states at the
Fermi level. 
This latter fact is consistent with the K-ELNES spectrum
\cite{ts.ku.05},
 which provides information about the DOS above the Fermi 
energy (cf. Fig.~\ref{doscomp}).
This suppression is not reproduced
by LDA electronic structure calculations, suggesting that strong
electronic correlations may play an important role for this material. 
In a previous work~\cite{so.ab.97}, it was suggested,
in order to improve the agreement with the measured spectra, 
 to convolute the
computed DOS with a combination of a Lorentzian and a Gaussian curve,
taking into account the broadening due to lifetime and experimental resolution effects. 
This treatment indeed improves on the overall shape producing a peak at an
energy of about $-1.5$ eV, although
 one should point out 
 that the fitted broadening parameters are much too large
(around $0.6 $ and $0.8$eV, respectively).

\begin{figure}[h]
\includegraphics[width=0.95\columnwidth]{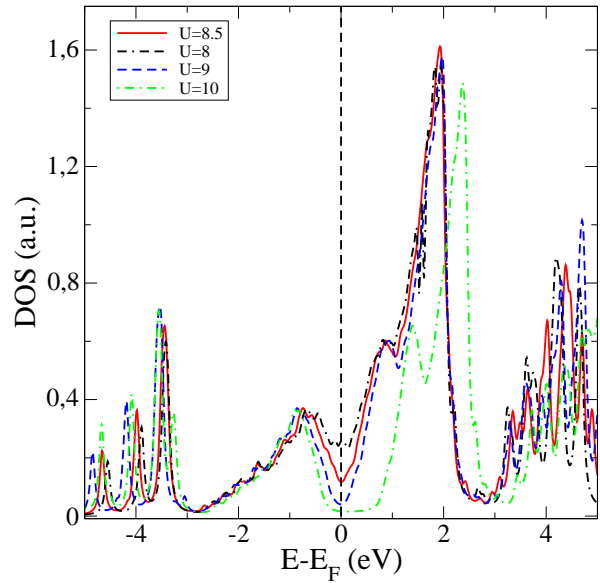} 
\caption{(color online) TiN density of states calculated with 
LDA+VCA for different values of the Coulomb-interaction 
parameter $U$ and
for $J=1.3$.}
\label{diffu}
\end{figure}

In the present paper, we argue that the pseudogap observed in the 
DOS of stoichiometric TiN 
is intrinsic to this material and signals the proximity to a
Mott metal-insulator transition.
To show this 
we start by calculating the DOS obtained from the Hamiltonian
\eqref{hi} with the value of $U=10$ eV obtained from constrained LDA.
The results, displayed in Fig.~\ref{diffu}, predict for this value of $U$
a Mott-insulating state with a gap of about $1 $eV. However, this is in
contrast to the experimental situation, since electrical conductivity
in TiN shows a metallic behavior, although with a relatively low residual 
conductivity~\cite{su.he.97}. Since results  obtained from constrained 
LDA are only approximate and tend to overestimate $U$ due to the delocalized 
nature of Wannier orbitals, we have also presented results  for  smaller
values of $U$ down to $8 $eV  (Fig.~\ref{diffu}). As one can see, 
no significant changes can be detected for states more than $3$eV away from the 
Fermi energy. Here, only static correlations, which get absorbed into the 
chemical potential, play a role. On the other hand, the situation changes 
rapidly around the Fermi energy. In particular, our results show that the 
Mott gap starts closing at $U\approx 9$eV, and at smaller $U$ down to 
$U\approx 8 $eV it acquires the shape of a pseudogap. As a matter of facts, 
the curve for $U=8.5$, shown in Fig.~\ref{doscomp} agrees quite well with 
experimental measurements. Notice that, in order to avoid introducing 
{\it ad-hoc } parameters, and to show fine-detailed features of the spectrum, 
the calculated curve has not been additionally broadened to meet experimental resolution. Taking this into
account, we see that our results reproduce quite well the experimental
features both above (K-ELNES) and below (XPS) the Fermi energy, and in particular the
``pseudogap'' behavior between $-1.0$ and $1.0$eV.
We stress that we are using an effective low-energy model which is
expected to correctly reproduce correlation effects close to the Fermi
energy, but is not expected to reproduce features beyond the range of
the NMTO bands shown in Fig.~\ref{NMTO_bnds} (full/red curves).
To reproduce the spectrum at higher binding energies, LDA and LDA+U
are more appropriate. In this sense, our results complement 
these techniques in the low-energy region.

Another remarkable feature of the computed LDA+VCA
 density of states spectra is the presence of a  set
of peaks in the energy range of $-5$ to $-3$eV. These states only appear 
when a sufficiently large cluster is taken as a reference system. Therefore, 
these states can be seen as non-local many-body incoherent features.
It is tempting to associate these states with the corresponding
``kink'' in the XPS spectrum~\cite{so.ab.97} at energies around $-4$eV.
However, this kink is also affected by the presence of N-$2p$ states whose 
bands start at $\approx -4 eV$ (see Fig.~\ref{doslda}). Therefore, our 
result suggests a strong hybridization between these states and the correlated 
many-body structures at $\approx -4 $eV.

The nitrogen K-ELNES spectrum is related to the unoccupied partial density of states 
with $p$-symmetry at the nitrogen site. The peaks situated at
energies around $2$ and     
 $4$eV can be attributed to the unoccupied N-$2p$ states hybridized  with Ti $3d$ states. 
Notice that the downfolded operators used in our effective Hamiltonian
\eqref{h0}+\eqref{hi} describe anyway an effective particle 
ultimately producing the bands of Fig.~\ref{NMTO_bnds}, i.e. 
the corresponding particle also contain a ``mixture'' of 
other orbitals such as N-$p$, in order to correctly reproduce 
the hybridization. It is remarkable to note that 
our calculation with $U=8.5$eV captures most of the features of the experimental 
spectra at both peaks, in particular the separation of the two peaks is in very 
good agreement with the experimental value $2.3$eV~\cite{pa.sc.00}.

It is important to mention that combining the experimental data from the occupied and unoccupied 
parts of the spectra with electric and magnetic properties of TiN
\cite{su.he.97}, a peculiar
metallic behavior emerges. Magnetic susceptibility measurements show that TiN is a Pauli-paramagnet,
and electrical conductivity demonstrates a metallic behavior with a
relatively large residual resistivity. 
In combination with the measured  
XPS spectra, one can conclude that at the Fermi level a pseudogap in
the density of states is formed, signaling the vicinity of a Mott-insulating phase.

The present LDA+VCA calculation includes correlation effects exactly on a length scale of an octahedral cluster consisting of 6
sites shown in Fig.~\ref{tin_fig} (connected by dashed lines) . Notable results of our present calculations are the correct 
description of the  $-1$eV peak and the double peak in the occupied part of the spectra. In addition,
at lower energies, many-body non-local incoherent features are formed which can explain the
kink in the XPS spectra situated at $-4$eV. Therefore, it is clear that the present LDA+VCA results
show notable improvements with respect to previous DFT-results in the
low-energy region, and explain the peculiar metallic behavior
of TiN.

A remark should be made concerning vacancy effects.
It is well known that  transition metal carbides and nitrides usually contain vacancies in
the metalloid (N) sublattice. The presence of vacancies  profoundly
influences the physical 
properties of this family of compounds. It is known from electronic-structure calculations 
that the presence of vacancies reduces the partial nitrogen $s$ and $p$ density of states, and 
produces additional peaks close to the Fermi level. 
These {\it vacancy peaks} 
show up in the  LDA-DOS of non-stoichiometric TiN
at about $-2$eV and in the vicinity of the Fermi level 
\cite{he.re.87,zh.gu.88,jh.lo.01,dr.bo.02}.
The peak at $-2 $eV  is associated with $\sigma$-bonding between Ti atoms through the N vacancy,
while the peak at $E_F$ is related to the $\sigma$-bonding between 
the nearest-neighbor Ti atoms
\cite{he.re.87,zh.gu.88}.
A quantitative comparison of the measured spectra and the broadened density of states 
obtained for non-stoichiometric materials
shows that the calculated density of states peaks are too narrow and shifted to lower 
binding energies with respect to experimental results.
The shift and the narrowing 
of the theoretical peaks was interpreted~\cite{re.ma.86} as due to 
limitations of the local density approximation
in describing electronic interactions.

\subsection{Spectral properties}
\label{spec}

\begin{figure}[h]
\includegraphics[width=0.99\columnwidth]{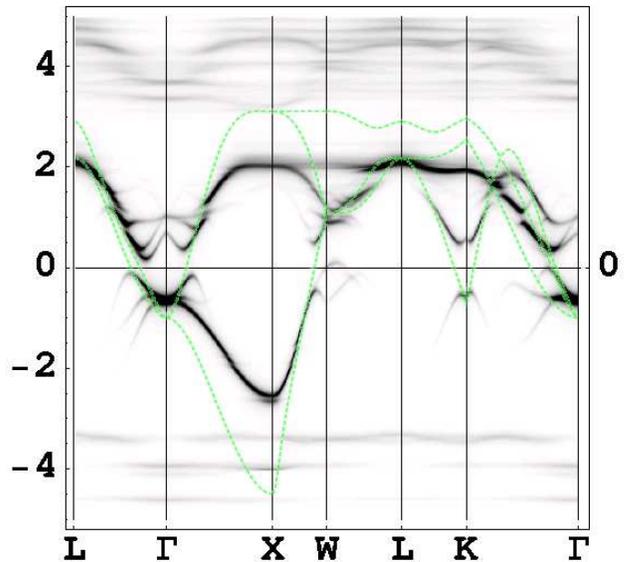}
\caption{Spectral-function of TiN for $U=8.5$ and $J=1.3$
 shown as density plot for 
the same path in the BZ as in Fig.~\ref{NMTO_bnds}. The bands obtained from downfolding (dashed, green lines) are shown as reference.}
\label{tin-spect}
\end{figure}

In order to gain insight into {\bf k}-dependent features of the local DOS, 
we have computed the {\bf k}-resolved spectral function $A({\bf k}, \omega)$,
which is plotted in Fig.~\ref{tin-spect}. 
From this figure one can clearly discern two metallic bands crossing the
Fermi energy near to the W point, and one between $K$ and $\Gamma$, 
however, with a small spectral
weight, consistent with the pseudogap picture discussed in Sec.~\ref{dos}.

Moreover, in the energy range between $-3$ to $-5$eV the spectral function in 
Fig.~\ref{tin-spect} shows dispersion-less features
 which are responsible 
for the non-local correlation peaks  discussed in the previous section. 
%
As discussed above, these features cannot be captured within 
a single-site LDA+Dynamical-Mean-Field (DMFT) approach, so that a cluster 
approach, as the one presented here, is required. Also the cluster geometry itself
is important: we chose an octahedral reference cluster (Fig.~\ref{tin_fig}) in
order to conserve the lattice symmetry. This is needed, as we have verified that
in smaller clusters, without the lattice symmetry, this feature is not present.

As discussed above, our calculations show that dynamical correlations
are important for TiN. On the other hand, it turns out that also {\it
  non-local } correlations
are crucial.
An important consequence of
this fact is that we do not expect the properties of TiN discussed here
to be
correctly reproduced by a single-site DMFT calculation.

\begin{figure}[h]
\includegraphics[width=0.99\columnwidth]{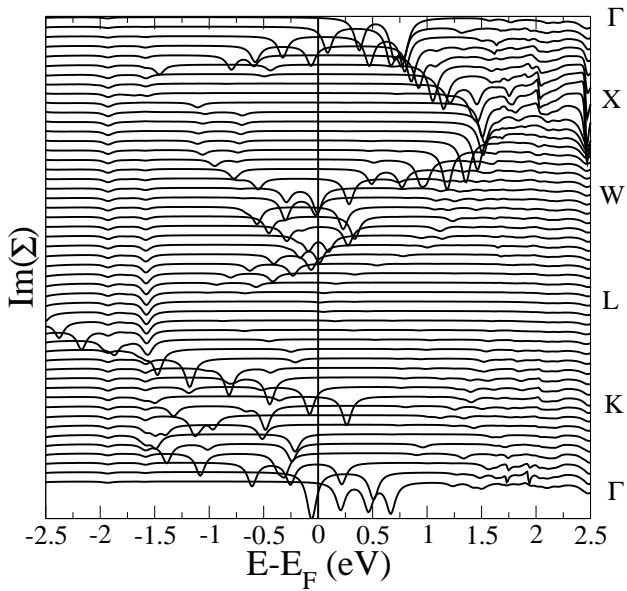}
\caption{Imaginary part of the self-energy 
calculated for the same parameters and
along the same BZ path as in Fig.~\ref{tin-spect}, and
presented in a three-dimensional plot.}
\label{sigma}
\end{figure}

To show this, we plot in Fig.~\ref{sigma}  the self-energy on the same
path around the BZ as for the spectral function (Fig.~\ref{tin-spect}).
As one can see, there is a strong {\bf k} dependence, especially in the
region around the Fermi energy. This
is an indication for the non-locality of the self-energy.
This non-locality strongly affects the metal-insulator transition as
well, since the most dispersive part of $\Sigma$ is precisely around the Fermi
energy.
Furthermore, there are contributions to the self-energy at larger
energies (not shown in Fig.~\ref{sigma}), which, however, are
essentially flat as a function of {\bf k}, and, thus, localized in
real space.

The {\bf k}-resolved spectral function allows us to determine the LDA+VCA quasiparticle 
bands, which are shown as density plot in Fig.~\ref{tin-spect} and are
compared with the  bands obtained from LDA (dashed). 
One can observe a reduction of the quasiparticle bandwidth (BW) from approximately $7.5$ eV in LDA to 
$4.5$ eV in LDA+VCA. Accordingly, the band narrowing due to correlation effects can be described in 
terms of a ``high-energy'' (HE) mass-renormalization factor $m_{HE}/m_{HE,LDA}\approx 1.7$.
Experimental measurements provide the total ratio
 $m_{HE}/m_e$ ($m_e$ is the free electron mass), which contains also
 the mass renormalization factor $m_{HE,LDA}/m_e$ coming from the band
 structure. 
The latter can  be roughly estimated 
by equating $\hbar^2 k_F^2/(2 m_{HE,LDA}) = E_F^*$, whereby 
$E_F^*=4.42 eV$ is the Ti-$3d$ band ``depth'', i. e. 
the Fermi energy measured from the bottom of the occupied
Ti-$3d$ band, and $k_F$ is related to the corresponding occupation $n=4/a^3$
by the usual relation $k_F^3=3 \pi^2 n$.
This gives $m_{HE,LDA}/m_e \approx 1.3$, producing an overall 
high-energy mass renormalization
$m_{HE}/m_e=2.2$.
This has to be distinguished from the low-energy  effective mass given
by the change in slope at the Fermi energy, i. e. the Fermi velocity.
Nevertheless, available experimental measurements discussed below provide
an estimate for $m_{HE}$ and not for the low-energy contribution.

Experimentally, an estimate of the electron effective mass in TiN 
has been obtained by UPS and EELS experiments using high-energy synchrotron radiation
\cite{wa.ma.98}. These methods also allow to determine the
band density $n$ of the $d$-electrons and the $3d$ band depth
(corresponding to our $E^*_F$ above). To estimate the
$d$-electron density several characteristics of the UPS spectrum
such as peak area, photoemission cross section  and
inelastic mean free path are used. The estimation of the
band depth
is obtained from the position of the minimum
in intensity between the valence and conduction bands~\cite{wa.ma.98}.
Considering the rough estimate we have used,
the computed
high-energy mass renormalization factor $m_{HE}/m_e=2.2$, turns out to
be in
 reasonable agreement with the value $2.7 \pm 0.3$,
obtained in Ref.~\onlinecite{wa.ma.98}, table 3, for the stoichiometric TiN
sample d.
%
%
%
%
%
%
%
%


\section{Summary}
\label{summ}

In this paper we have analyzed the physical properties of TiN in a combined electronic-structure 
and many-body approach. The NMTO downfolding technique was
 used to calculate the 
LDA low-energy effective Hamiltonian for the $Ti-d(t_{2g})$ bands. 
The effect of non-local
correlations was treated within a variational cluster perturbation approach, which deals
 exactly with correlations on the cluster scale. We present results for the 
local density of states, the
self-energy, the spectral function and for the effective mass.
The results are analyzed in comparison with experiments on
valence band XPS and K-ELNES spectra and, in the case of the effective mass, compared to UPS and EELS experiments. 

Our results 
suggest that TiN is a 
peculiar metal with a pseudogap at the Fermi level indicating the
proximity to a metal-insulator transition.
In our calculations the pseudogap regime is best described for a value of $U=8.5$eV
for the Coulomb-interaction. 

It is important to mention that 
this result could be achieved only by an appropriate treatment of
dynamical and non-local correlations. In particular, neither LDA nor
LDA+U calculations provide even a qualitative description of the DOS
suppression at the Fermi energy. 
Due to the non-locality of the self-energy, we expect that even
single-site DMFT may not be appropriate for a proper description of this system
at low energies.

Our calculations provide a
 good qualitative and semiquantitative agreement with experiments,
when comparing the DOS with XPS and K-ELNES spectra close to the Fermi
energy.
In particular, our  results show a coherent feature situated at $-1$eV 
which can be clearly identified in the XPS spectra. We stress that none of previous theoretical 
investigations are able to capture this particular low energy feature. Furthermore, at low
energies our results provide a qualitative interpretation of  the
$-4$eV kink in the XPS spectra as an hybridization of incoherent non-local 
many-body features with the N-$p$ states. 
Also, in the unoccupied part above the Fermi level, our results 
reproduce the double peak structure in good agreement with the K-ELNES
spectra. 
Finally, we calculated the effective mass from high- and low-energy spectral features and 
found reasonable agreement with experimental values.

\section*{Acknowledgments}

We are grateful to Karsten Held for helpful suggestions.
This work was supported
by the Austrian science fund (FWF project P18505-N16), and by the 
cooperation project ``NAWI Graz''  
(F-NW-515-GASS).  L.C. also acknowledges the financial
support offered by Romanian Grant CNCSIS/ID672/2009.

\bibliography{references_database}

\begin{thebibliography}{46}
\expandafter\ifx\csname natexlab\endcsname\relax\def\natexlab#1{#1}\fi
\expandafter\ifx\csname bibnamefont\endcsname\relax
  \def\bibnamefont#1{#1}\fi
\expandafter\ifx\csname bibfnamefont\endcsname\relax
  \def\bibfnamefont#1{#1}\fi
\expandafter\ifx\csname citenamefont\endcsname\relax
  \def\citenamefont#1{#1}\fi
\expandafter\ifx\csname url\endcsname\relax
  \def\url#1{\texttt{#1}}\fi
\expandafter\ifx\csname urlprefix\endcsname\relax\def\urlprefix{URL }\fi
\providecommand{\bibinfo}[2]{#2}
\providecommand{\eprint}[2][]{\url{#2}}

\bibitem[{\citenamefont{Schwarz et~al.}(1985)\citenamefont{Schwarz, Williams,
  Cuomo, Harper, and Hentzell}}]{sc.wi.85}
\bibinfo{author}{\bibfnamefont{K.}~\bibnamefont{Schwarz}},
  \bibinfo{author}{\bibfnamefont{A.~R.} \bibnamefont{Williams}},
  \bibinfo{author}{\bibfnamefont{J.~J.} \bibnamefont{Cuomo}},
  \bibinfo{author}{\bibfnamefont{J.~H.~E.} \bibnamefont{Harper}},
  \bibnamefont{and} \bibinfo{author}{\bibfnamefont{H.~T.~G.}
  \bibnamefont{Hentzell}}, \bibinfo{journal}{Phys. Rev. B}
  \textbf{\bibinfo{volume}{32}}, \bibinfo{pages}{8312} (\bibinfo{year}{1985}).

\bibitem[{\citenamefont{Vinokur et~al.}(2008)\citenamefont{Vinokur, Baturina,
  Fistul, Mironov, Baklanov, and Strunk}}]{vi.ba.08}
\bibinfo{author}{\bibfnamefont{V.~M.} \bibnamefont{Vinokur}},
  \bibinfo{author}{\bibfnamefont{T.~I.} \bibnamefont{Baturina}},
  \bibinfo{author}{\bibfnamefont{M.~V.} \bibnamefont{Fistul}},
  \bibinfo{author}{\bibfnamefont{A.~Y.} \bibnamefont{Mironov}},
  \bibinfo{author}{\bibfnamefont{M.~R.} \bibnamefont{Baklanov}},
  \bibnamefont{and} \bibinfo{author}{\bibfnamefont{C.}~\bibnamefont{Strunk}},
  \bibinfo{journal}{Nature} \textbf{\bibinfo{volume}{452}},
  \bibinfo{pages}{613} (\bibinfo{year}{2008}).

\bibitem[{\citenamefont{Soriano et~al.}(1997)\citenamefont{Soriano, Abbate,
  Pen, Prieto, and Sanz}}]{so.ab.97}
\bibinfo{author}{\bibfnamefont{L.}~\bibnamefont{Soriano}},
  \bibinfo{author}{\bibfnamefont{M.}~\bibnamefont{Abbate}},
  \bibinfo{author}{\bibfnamefont{H.}~\bibnamefont{Pen}},
  \bibinfo{author}{\bibfnamefont{P.}~\bibnamefont{Prieto}}, \bibnamefont{and}
  \bibinfo{author}{\bibfnamefont{J.~M.} \bibnamefont{Sanz}},
  \bibinfo{journal}{Solid State Communications} \textbf{\bibinfo{volume}{102}},
  \bibinfo{pages}{291 } (\bibinfo{year}{1997}).

\bibitem[{\citenamefont{Porte et~al.}(1983)\citenamefont{Porte, Roux, and
  Hanus}}]{po.ro.83}
\bibinfo{author}{\bibfnamefont{L.}~\bibnamefont{Porte}},
  \bibinfo{author}{\bibfnamefont{L.}~\bibnamefont{Roux}}, \bibnamefont{and}
  \bibinfo{author}{\bibfnamefont{J.}~\bibnamefont{Hanus}},
  \bibinfo{journal}{Phys. Rev. B} \textbf{\bibinfo{volume}{28}},
  \bibinfo{pages}{3214} (\bibinfo{year}{1983}).

\bibitem[{\citenamefont{Johansson et~al.}(1980)\citenamefont{Johansson, Stefan,
  Shek, and Christensen}}]{jo.st.80}
\bibinfo{author}{\bibfnamefont{L.~I.} \bibnamefont{Johansson}},
  \bibinfo{author}{\bibfnamefont{P.~M.} \bibnamefont{Stefan}},
  \bibinfo{author}{\bibfnamefont{M.~L.} \bibnamefont{Shek}}, \bibnamefont{and}
  \bibinfo{author}{\bibfnamefont{A.~N.} \bibnamefont{Christensen}},
  \bibinfo{journal}{Phys. Rev. B} \textbf{\bibinfo{volume}{22}},
  \bibinfo{pages}{1032} (\bibinfo{year}{1980}).

\bibitem[{\citenamefont{Gubanov et~al.}(1977)\citenamefont{Gubanov, Kurmaev,
  and Shveikin}}]{gu.ku.77}
\bibinfo{author}{\bibfnamefont{V.~A.} \bibnamefont{Gubanov}},
  \bibinfo{author}{\bibfnamefont{E.~Z.} \bibnamefont{Kurmaev}},
  \bibnamefont{and} \bibinfo{author}{\bibfnamefont{G.~P.}
  \bibnamefont{Shveikin}}, \bibinfo{journal}{J. Phys. Chem. Solids}
  \textbf{\bibinfo{volume}{38}}, \bibinfo{pages}{201} (\bibinfo{year}{1977}).

\bibitem[{\citenamefont{Riehle et~al.}(1982)\citenamefont{Riehle, Wolf, and
  Politis}}]{ri.wo.82}
\bibinfo{author}{\bibfnamefont{F.}~\bibnamefont{Riehle}},
  \bibinfo{author}{\bibfnamefont{T.}~\bibnamefont{Wolf}}, \bibnamefont{and}
  \bibinfo{author}{\bibfnamefont{C.}~\bibnamefont{Politis}},
  \bibinfo{journal}{Z. Phys. B} \textbf{\bibinfo{volume}{47}},
  \bibinfo{pages}{3} (\bibinfo{year}{1982}).

\bibitem[{\citenamefont{Schubert
  et~al.}(1981{\natexlab{a}})\citenamefont{Schubert, Shelton, and
  Wolf}}]{sc.sh.81}
\bibinfo{author}{\bibfnamefont{W.~K.} \bibnamefont{Schubert}},
  \bibinfo{author}{\bibfnamefont{R.~N.} \bibnamefont{Shelton}},
  \bibnamefont{and} \bibinfo{author}{\bibfnamefont{E.~L.} \bibnamefont{Wolf}},
  \bibinfo{journal}{Phys. Rev. B} \textbf{\bibinfo{volume}{23}},
  \bibinfo{pages}{5097} (\bibinfo{year}{1981}{\natexlab{a}}).

\bibitem[{\citenamefont{Schubert
  et~al.}(1981{\natexlab{b}})\citenamefont{Schubert, Shelton, and
  Wolf}}]{sc.sh.81b}
\bibinfo{author}{\bibfnamefont{W.~K.} \bibnamefont{Schubert}},
  \bibinfo{author}{\bibfnamefont{R.~N.} \bibnamefont{Shelton}},
  \bibnamefont{and} \bibinfo{author}{\bibfnamefont{E.~L.} \bibnamefont{Wolf}},
  \bibinfo{journal}{Phys. Rev. B} \textbf{\bibinfo{volume}{24}},
  \bibinfo{pages}{6278} (\bibinfo{year}{1981}{\natexlab{b}}).

\bibitem[{\citenamefont{Lynch et~al.}(1980)\citenamefont{Lynch, Olson,
  Peterman, and Weaver}}]{ly.ol.80}
\bibinfo{author}{\bibfnamefont{D.~W.} \bibnamefont{Lynch}},
  \bibinfo{author}{\bibfnamefont{C.~G.} \bibnamefont{Olson}},
  \bibinfo{author}{\bibfnamefont{D.~J.} \bibnamefont{Peterman}},
  \bibnamefont{and} \bibinfo{author}{\bibfnamefont{J.~H.}
  \bibnamefont{Weaver}}, \bibinfo{journal}{Phys. Rev. B}
  \textbf{\bibinfo{volume}{22}}, \bibinfo{pages}{3991} (\bibinfo{year}{1980}).

\bibitem[{\citenamefont{Alward et~al.}(1975)\citenamefont{Alward, Fong,
  El-Batanouny, and Wooten}}]{al.fo.75}
\bibinfo{author}{\bibfnamefont{J.~F.} \bibnamefont{Alward}},
  \bibinfo{author}{\bibfnamefont{C.~Y.} \bibnamefont{Fong}},
  \bibinfo{author}{\bibfnamefont{M.}~\bibnamefont{El-Batanouny}},
  \bibnamefont{and} \bibinfo{author}{\bibfnamefont{F.}~\bibnamefont{Wooten}},
  \bibinfo{journal}{Phys. Rev. B} \textbf{\bibinfo{volume}{12}},
  \bibinfo{pages}{1105} (\bibinfo{year}{1975}).

\bibitem[{\citenamefont{Neckel}(1983)}]{neck.83}
\bibinfo{author}{\bibfnamefont{A.}~\bibnamefont{Neckel}},
  \bibinfo{journal}{International Journal of Quantum Chemistry}
  \textbf{\bibinfo{volume}{13}}, \bibinfo{pages}{1317} (\bibinfo{year}{1983}).

\bibitem[{\citenamefont{Marksteiner et~al.}(1986)\citenamefont{Marksteiner,
  Weinberger, Neckel, Zeller, and Dederichs}}]{ma.we.86}
\bibinfo{author}{\bibfnamefont{P.}~\bibnamefont{Marksteiner}},
  \bibinfo{author}{\bibfnamefont{P.}~\bibnamefont{Weinberger}},
  \bibinfo{author}{\bibfnamefont{A.}~\bibnamefont{Neckel}},
  \bibinfo{author}{\bibfnamefont{R.}~\bibnamefont{Zeller}}, \bibnamefont{and}
  \bibinfo{author}{\bibfnamefont{P.~H.} \bibnamefont{Dederichs}},
  \bibinfo{journal}{Phys. Rev. B} \textbf{\bibinfo{volume}{33}},
  \bibinfo{pages}{812} (\bibinfo{year}{1986}).

\bibitem[{\citenamefont{Papaconstantopoulos
  et~al.}(1985)\citenamefont{Papaconstantopoulos, Pickett, Klein, and
  Boyer}}]{pa.pi.85}
\bibinfo{author}{\bibfnamefont{D.~A.} \bibnamefont{Papaconstantopoulos}},
  \bibinfo{author}{\bibfnamefont{W.~E.} \bibnamefont{Pickett}},
  \bibinfo{author}{\bibfnamefont{B.~M.} \bibnamefont{Klein}}, \bibnamefont{and}
  \bibinfo{author}{\bibfnamefont{L.~L.} \bibnamefont{Boyer}},
  \bibinfo{journal}{Phys. Rev. B} \textbf{\bibinfo{volume}{31}},
  \bibinfo{pages}{752} (\bibinfo{year}{1985}).

\bibitem[{\citenamefont{Zhukov et~al.}(1988)\citenamefont{Zhukov, Gubanov,
  Jepsen, Christensen, and Andersen}}]{zh.gu.88}
\bibinfo{author}{\bibfnamefont{V.~P.} \bibnamefont{Zhukov}},
  \bibinfo{author}{\bibfnamefont{V.~A.} \bibnamefont{Gubanov}},
  \bibinfo{author}{\bibfnamefont{O.}~\bibnamefont{Jepsen}},
  \bibinfo{author}{\bibfnamefont{N.~E.} \bibnamefont{Christensen}},
  \bibnamefont{and} \bibinfo{author}{\bibfnamefont{O.~K.}
  \bibnamefont{Andersen}}, \bibinfo{journal}{J. Phys. Chem. Solids}
  \textbf{\bibinfo{volume}{49}}, \bibinfo{pages}{841} (\bibinfo{year}{1988}).

\bibitem[{\citenamefont{Jhi et~al.}(2001)\citenamefont{Jhi, Louie, Cohen, and
  Ihm}}]{jh.lo.01}
\bibinfo{author}{\bibfnamefont{S.-H.} \bibnamefont{Jhi}},
  \bibinfo{author}{\bibfnamefont{S.~G.} \bibnamefont{Louie}},
  \bibinfo{author}{\bibfnamefont{M.~L.} \bibnamefont{Cohen}}, \bibnamefont{and}
  \bibinfo{author}{\bibfnamefont{J.}~\bibnamefont{Ihm}},
  \bibinfo{journal}{Phys. Rev. Lett.} \textbf{\bibinfo{volume}{86}},
  \bibinfo{pages}{3348} (\bibinfo{year}{2001}).

\bibitem[{\citenamefont{Paxton et~al.}(2000)\citenamefont{Paxton, van
  Schilfgaarde, MacKenzie, and Craven}}]{pa.sc.00}
\bibinfo{author}{\bibfnamefont{A.~T.} \bibnamefont{Paxton}},
  \bibinfo{author}{\bibfnamefont{M.}~\bibnamefont{van Schilfgaarde}},
  \bibinfo{author}{\bibfnamefont{M.}~\bibnamefont{MacKenzie}},
  \bibnamefont{and} \bibinfo{author}{\bibfnamefont{A.~J.}
  \bibnamefont{Craven}}, \bibinfo{journal}{J. Phys.: Condens. Matter}
  \textbf{\bibinfo{volume}{12}}, \bibinfo{pages}{729} (\bibinfo{year}{2000}).

\bibitem[{\citenamefont{Anisimov et~al.}(1997)\citenamefont{Anisimov,
  Aryasetiawan, and Lichtenstein}}]{an.ar.97}
\bibinfo{author}{\bibfnamefont{V.~I.} \bibnamefont{Anisimov}},
  \bibinfo{author}{\bibfnamefont{F.}~\bibnamefont{Aryasetiawan}},
  \bibnamefont{and} \bibinfo{author}{\bibfnamefont{A.~I.}
  \bibnamefont{Lichtenstein}}, \bibinfo{journal}{Journal of Physics: Condensed
  Matter} \textbf{\bibinfo{volume}{9}}, \bibinfo{pages}{767}
  (\bibinfo{year}{1997}).

\bibitem[{\citenamefont{Potthoff et~al.}(2003)\citenamefont{Potthoff, Aichhorn,
  and Dahnken}}]{po.ai.03}
\bibinfo{author}{\bibfnamefont{M.}~\bibnamefont{Potthoff}},
  \bibinfo{author}{\bibfnamefont{M.}~\bibnamefont{Aichhorn}}, \bibnamefont{and}
  \bibinfo{author}{\bibfnamefont{C.}~\bibnamefont{Dahnken}},
  \bibinfo{journal}{Phys. Rev. Lett.} \textbf{\bibinfo{volume}{91}},
  \bibinfo{pages}{206402} (\bibinfo{year}{2003}).

\bibitem[{\citenamefont{Dahnken et~al.}(2004)\citenamefont{Dahnken, Aichhorn,
  Hanke, Arrigoni, and Potthoff}}]{da.ai.04}
\bibinfo{author}{\bibfnamefont{C.}~\bibnamefont{Dahnken}},
  \bibinfo{author}{\bibfnamefont{M.}~\bibnamefont{Aichhorn}},
  \bibinfo{author}{\bibfnamefont{W.}~\bibnamefont{Hanke}},
  \bibinfo{author}{\bibfnamefont{E.}~\bibnamefont{Arrigoni}}, \bibnamefont{and}
  \bibinfo{author}{\bibfnamefont{M.}~\bibnamefont{Potthoff}},
  \bibinfo{journal}{Phys. Rev. B} \textbf{\bibinfo{volume}{70}},
  \bibinfo{pages}{245110} (\bibinfo{year}{2004}).

\bibitem[{\citenamefont{Andersen and Jepsen}(1984)}]{an.je.84}
\bibinfo{author}{\bibfnamefont{O.~K.} \bibnamefont{Andersen}} \bibnamefont{and}
  \bibinfo{author}{\bibfnamefont{O.}~\bibnamefont{Jepsen}},
  \bibinfo{journal}{Phys. Rev. Lett.} \textbf{\bibinfo{volume}{53}},
  \bibinfo{pages}{2571} (\bibinfo{year}{1984}).

\bibitem[{\citenamefont{Andersen and Saha-Dasgupta}(2000)}]{an.sa.00}
\bibinfo{author}{\bibfnamefont{O.~K.} \bibnamefont{Andersen}} \bibnamefont{and}
  \bibinfo{author}{\bibfnamefont{T.}~\bibnamefont{Saha-Dasgupta}},
  \bibinfo{journal}{Phys. Rev. B} \textbf{\bibinfo{volume}{62}},
  \bibinfo{pages}{R16219} (\bibinfo{year}{2000}).

\bibitem[{\citenamefont{Zurek et~al.}(2005)\citenamefont{Zurek, Jepsen, and
  Andersen}}]{zu.je.05}
\bibinfo{author}{\bibfnamefont{E.}~\bibnamefont{Zurek}},
  \bibinfo{author}{\bibfnamefont{O.}~\bibnamefont{Jepsen}}, \bibnamefont{and}
  \bibinfo{author}{\bibfnamefont{O.~K.} \bibnamefont{Andersen}},
  \bibinfo{journal}{ChemPhysChem} \textbf{\bibinfo{volume}{6}},
  \bibinfo{pages}{1934} (\bibinfo{year}{2005}).

\bibitem[{\citenamefont{Anisimov and Gunnarsson}(1991)}]{an.gu.91}
\bibinfo{author}{\bibfnamefont{V.~I.} \bibnamefont{Anisimov}} \bibnamefont{and}
  \bibinfo{author}{\bibfnamefont{O.}~\bibnamefont{Gunnarsson}},
  \bibinfo{journal}{Phys. Rev. B} \textbf{\bibinfo{volume}{43}},
  \bibinfo{pages}{7570} (\bibinfo{year}{1991}).

\bibitem[{\citenamefont{Anisimov et~al.}(1991)\citenamefont{Anisimov, Zaanen,
  and Andersen}}]{an.za.91}
\bibinfo{author}{\bibfnamefont{V.~I.} \bibnamefont{Anisimov}},
  \bibinfo{author}{\bibfnamefont{J.}~\bibnamefont{Zaanen}}, \bibnamefont{and}
  \bibinfo{author}{\bibfnamefont{O.~K.} \bibnamefont{Andersen}},
  \bibinfo{journal}{Phys. Rev. B} \textbf{\bibinfo{volume}{44}},
  \bibinfo{pages}{943} (\bibinfo{year}{1991}).

\bibitem[{\citenamefont{Pavarini et~al.}(2005)\citenamefont{Pavarini, Yamasaki,
  Nuss, and Andersen}}]{pa.ya.05}
\bibinfo{author}{\bibfnamefont{E.}~\bibnamefont{Pavarini}},
  \bibinfo{author}{\bibfnamefont{A.}~\bibnamefont{Yamasaki}},
  \bibinfo{author}{\bibfnamefont{J.}~\bibnamefont{Nuss}}, \bibnamefont{and}
  \bibinfo{author}{\bibfnamefont{O.~K.} \bibnamefont{Andersen}},
  \bibinfo{journal}{New J. Phys.} \textbf{\bibinfo{volume}{7}},
  \bibinfo{pages}{188} (\bibinfo{year}{2005}).

\bibitem[{\citenamefont{Aryasetiawan et~al.}(2006)\citenamefont{Aryasetiawan,
  Karlsson, Jepsen, and Sch\"onberger}}]{ar.ka.06}
\bibinfo{author}{\bibfnamefont{F.}~\bibnamefont{Aryasetiawan}},
  \bibinfo{author}{\bibfnamefont{K.}~\bibnamefont{Karlsson}},
  \bibinfo{author}{\bibfnamefont{O.}~\bibnamefont{Jepsen}}, \bibnamefont{and}
  \bibinfo{author}{\bibfnamefont{U.}~\bibnamefont{Sch\"onberger}},
  \bibinfo{journal}{Phys. Rev. B} \textbf{\bibinfo{volume}{74}},
  \bibinfo{pages}{125106} (\bibinfo{year}{2006}).

\bibitem[{\citenamefont{Aryasetiawan et~al.}(2004)\citenamefont{Aryasetiawan,
  Imada, Georges, Kotliar, Biermann, and Lichtenstein}}]{ar.im.04}
\bibinfo{author}{\bibfnamefont{F.}~\bibnamefont{Aryasetiawan}},
  \bibinfo{author}{\bibfnamefont{M.}~\bibnamefont{Imada}},
  \bibinfo{author}{\bibfnamefont{A.}~\bibnamefont{Georges}},
  \bibinfo{author}{\bibfnamefont{G.}~\bibnamefont{Kotliar}},
  \bibinfo{author}{\bibfnamefont{S.}~\bibnamefont{Biermann}}, \bibnamefont{and}
  \bibinfo{author}{\bibfnamefont{A.~I.} \bibnamefont{Lichtenstein}},
  \bibinfo{journal}{Phys. Rev. B} \textbf{\bibinfo{volume}{70}},
  \bibinfo{pages}{195104} (\bibinfo{year}{2004}).

\bibitem[{\citenamefont{Czyzyk and Sawatzky}(1994)}]{cz.sa.94}
\bibinfo{author}{\bibfnamefont{M.~T.} \bibnamefont{Czyzyk}} \bibnamefont{and}
  \bibinfo{author}{\bibfnamefont{G.~A.} \bibnamefont{Sawatzky}},
  \bibinfo{journal}{Phys. Rev. B} \textbf{\bibinfo{volume}{49}},
  \bibinfo{pages}{14211} (\bibinfo{year}{1994}).

\bibitem[{\citenamefont{Petukhov et~al.}(2003)\citenamefont{Petukhov, Mazin,
  Chioncel, and Lichtenstein}}]{pe.ma.03}
\bibinfo{author}{\bibfnamefont{A.~G.} \bibnamefont{Petukhov}},
  \bibinfo{author}{\bibfnamefont{I.~I.} \bibnamefont{Mazin}},
  \bibinfo{author}{\bibfnamefont{L.}~\bibnamefont{Chioncel}}, \bibnamefont{and}
  \bibinfo{author}{\bibfnamefont{A.~I.} \bibnamefont{Lichtenstein}},
  \bibinfo{journal}{Phys. Rev. B} \textbf{\bibinfo{volume}{67}},
  \bibinfo{pages}{153106} (\bibinfo{year}{2003}).

\bibitem[{\citenamefont{Gros and Valenti}(1993)}]{gr.va.93}
\bibinfo{author}{\bibfnamefont{C.}~\bibnamefont{Gros}} \bibnamefont{and}
  \bibinfo{author}{\bibfnamefont{R.}~\bibnamefont{Valenti}},
  \bibinfo{journal}{Phys. Rev. B} \textbf{\bibinfo{volume}{48}},
  \bibinfo{pages}{418} (\bibinfo{year}{1993}).

\bibitem[{\citenamefont{S{\'e}n{\'e}chal
  et~al.}(2000)\citenamefont{S{\'e}n{\'e}chal, Perez, and
  Pioro-Ladriere}}]{se.pe.00}
\bibinfo{author}{\bibfnamefont{D.}~\bibnamefont{S{\'e}n{\'e}chal}},
  \bibinfo{author}{\bibfnamefont{D.}~\bibnamefont{Perez}}, \bibnamefont{and}
  \bibinfo{author}{\bibfnamefont{M.}~\bibnamefont{Pioro-Ladriere}},
  \bibinfo{journal}{Phys. Rev. Lett.} \textbf{\bibinfo{volume}{84}},
  \bibinfo{pages}{522} (\bibinfo{year}{2000}).

\bibitem[{\citenamefont{Ovchinnikov and Sandalov}(1989)}]{ov.sa.89}
\bibinfo{author}{\bibfnamefont{S.~G.} \bibnamefont{Ovchinnikov}}
  \bibnamefont{and} \bibinfo{author}{\bibfnamefont{I.~S.}
  \bibnamefont{Sandalov}}, \bibinfo{journal}{Physica C}
  \textbf{\bibinfo{volume}{161}}, \bibinfo{pages}{607} (\bibinfo{year}{1989}).

\bibitem[{\citenamefont{Potthoff}(2003{\natexlab{a}})}]{pott.03}
\bibinfo{author}{\bibfnamefont{M.}~\bibnamefont{Potthoff}},
  \bibinfo{journal}{Eur. Phys. J. B} \textbf{\bibinfo{volume}{32}},
  \bibinfo{pages}{429} (\bibinfo{year}{2003}{\natexlab{a}}).

\bibitem[{\citenamefont{Potthoff}(2003{\natexlab{b}})}]{pott.03.se}
\bibinfo{author}{\bibfnamefont{M.}~\bibnamefont{Potthoff}},
  \bibinfo{journal}{Eur. Phys. J. B} \textbf{\bibinfo{volume}{36}},
  \bibinfo{pages}{335} (\bibinfo{year}{2003}{\natexlab{b}}).

\bibitem[{\citenamefont{Aichhorn and Arrigoni}(2005)}]{ai.ar.05}
\bibinfo{author}{\bibfnamefont{M.}~\bibnamefont{Aichhorn}} \bibnamefont{and}
  \bibinfo{author}{\bibfnamefont{E.}~\bibnamefont{Arrigoni}},
  \bibinfo{journal}{Europhys. Lett.} \textbf{\bibinfo{volume}{72}},
  \bibinfo{pages}{117} (\bibinfo{year}{2005}).

\bibitem[{\citenamefont{Aichhorn et~al.}(2006)\citenamefont{Aichhorn, Arrigoni,
  Potthoff, and Hanke}}]{ai.ar.06}
\bibinfo{author}{\bibfnamefont{M.}~\bibnamefont{Aichhorn}},
  \bibinfo{author}{\bibfnamefont{E.}~\bibnamefont{Arrigoni}},
  \bibinfo{author}{\bibfnamefont{M.}~\bibnamefont{Potthoff}}, \bibnamefont{and}
  \bibinfo{author}{\bibfnamefont{W.}~\bibnamefont{Hanke}},
  \bibinfo{journal}{Phys. Rev. B} \textbf{\bibinfo{volume}{74}},
  \bibinfo{pages}{024508} (\bibinfo{year}{2006}).

\bibitem[{\citenamefont{Lu and Arrigoni}(2009)}]{lu.ar.09u}
\bibinfo{author}{\bibfnamefont{X.}~\bibnamefont{Lu}} \bibnamefont{and}
  \bibinfo{author}{\bibfnamefont{E.}~\bibnamefont{Arrigoni}}
  (\bibinfo{year}{2009}), \bibinfo{note}{arXiv:0902.0388}.

\bibitem[{\citenamefont{Biroli et~al.}(2004)\citenamefont{Biroli, Parcollet,
  and Kotliar}}]{bi.pa.04}
\bibinfo{author}{\bibfnamefont{G.}~\bibnamefont{Biroli}},
  \bibinfo{author}{\bibfnamefont{O.}~\bibnamefont{Parcollet}},
  \bibnamefont{and} \bibinfo{author}{\bibfnamefont{G.}~\bibnamefont{Kotliar}},
  \bibinfo{journal}{Phys. Rev. B} \textbf{\bibinfo{volume}{69}},
  \bibinfo{pages}{205108} (\bibinfo{year}{2004}).

\bibitem[{\citenamefont{S{\'en\'echal}}(2008)}]{sene.08u}
\bibinfo{author}{\bibfnamefont{D.}~\bibnamefont{S{\'en\'echal}}}
  (\bibinfo{year}{2008}), \bibinfo{note}{arXiv:0806.2690}.

\bibitem[{\citenamefont{Tsujimoto et~al.}(2005)\citenamefont{Tsujimoto, Kurata,
  Nemoto, Isoda, Terada, and Kaji}}]{ts.ku.05}
\bibinfo{author}{\bibfnamefont{M.}~\bibnamefont{Tsujimoto}},
  \bibinfo{author}{\bibfnamefont{H.}~\bibnamefont{Kurata}},
  \bibinfo{author}{\bibfnamefont{T.}~\bibnamefont{Nemoto}},
  \bibinfo{author}{\bibfnamefont{S.}~\bibnamefont{Isoda}},
  \bibinfo{author}{\bibfnamefont{S.}~\bibnamefont{Terada}}, \bibnamefont{and}
  \bibinfo{author}{\bibfnamefont{K.}~\bibnamefont{Kaji}},
  \bibinfo{journal}{Journal of Electron Spectroscopy and Related Phenomena}
  \textbf{\bibinfo{volume}{143}}, \bibinfo{pages}{161 } (\bibinfo{year}{2005}).

\bibitem[{\citenamefont{Herle et~al.}(1997)\citenamefont{Herle, Hegde,
  Vasathacharyan, and Philip}}]{su.he.97}
\bibinfo{author}{\bibfnamefont{P.~S.} \bibnamefont{Herle}},
  \bibinfo{author}{\bibfnamefont{M.~S.} \bibnamefont{Hegde}},
  \bibinfo{author}{\bibfnamefont{N.~Y.} \bibnamefont{Vasathacharyan}},
  \bibnamefont{and} \bibinfo{author}{\bibfnamefont{S.}~\bibnamefont{Philip}},
  \bibinfo{journal}{J. Solid. Stat. Chem.} \textbf{\bibinfo{volume}{134}},
  \bibinfo{pages}{120} (\bibinfo{year}{1997}).

\bibitem[{\citenamefont{Herzig et~al.}(1987)\citenamefont{Herzig, Redinger,
  Eibler, and Neckel}}]{he.re.87}
\bibinfo{author}{\bibfnamefont{P.}~\bibnamefont{Herzig}},
  \bibinfo{author}{\bibfnamefont{J.}~\bibnamefont{Redinger}},
  \bibinfo{author}{\bibfnamefont{R.}~\bibnamefont{Eibler}}, \bibnamefont{and}
  \bibinfo{author}{\bibfnamefont{A.}~\bibnamefont{Neckel}},
  \bibinfo{journal}{J. Solid. Stat. Chem.} \textbf{\bibinfo{volume}{70}},
  \bibinfo{pages}{281} (\bibinfo{year}{1987}).

\bibitem[{\citenamefont{Dridi et~al.}(2002)\citenamefont{Dridi, Bouhafs,
  Ruterana, and Aourag}}]{dr.bo.02}
\bibinfo{author}{\bibfnamefont{Z.}~\bibnamefont{Dridi}},
  \bibinfo{author}{\bibfnamefont{B.}~\bibnamefont{Bouhafs}},
  \bibinfo{author}{\bibfnamefont{P.}~\bibnamefont{Ruterana}}, \bibnamefont{and}
  \bibinfo{author}{\bibfnamefont{H.}~\bibnamefont{Aourag}},
  \bibinfo{journal}{J. Phys.: Condens. Matter} \textbf{\bibinfo{volume}{14}},
  \bibinfo{pages}{10237} (\bibinfo{year}{2002}).

\bibitem[{\citenamefont{Redinger et~al.}(1986)\citenamefont{Redinger,
  Marksteiner, and Weinberger}}]{re.ma.86}
\bibinfo{author}{\bibfnamefont{J.}~\bibnamefont{Redinger}},
  \bibinfo{author}{\bibfnamefont{P.}~\bibnamefont{Marksteiner}},
  \bibnamefont{and}
  \bibinfo{author}{\bibfnamefont{P.}~\bibnamefont{Weinberger}},
  \bibinfo{journal}{Z. Phys. B} \textbf{\bibinfo{volume}{63}},
  \bibinfo{pages}{321} (\bibinfo{year}{1986}).

\bibitem[{\citenamefont{Walker et~al.}(1998)\citenamefont{Walker, Matthew,
  Anderson, and Brown}}]{wa.ma.98}
\bibinfo{author}{\bibfnamefont{C.~G.~H.} \bibnamefont{Walker}},
  \bibinfo{author}{\bibfnamefont{J.~A.~D.} \bibnamefont{Matthew}},
  \bibinfo{author}{\bibfnamefont{C.~A.} \bibnamefont{Anderson}},
  \bibnamefont{and} \bibinfo{author}{\bibfnamefont{N.~M.~D.}
  \bibnamefont{Brown}}, \bibinfo{journal}{Surf. Sci.}
  \textbf{\bibinfo{volume}{412/413}}, \bibinfo{pages}{405}
  (\bibinfo{year}{1998}).

\end{thebibliography}

\end{document}